\documentclass{aastex63}

\usepackage{amsmath} 
\usepackage{CJK} 
\usepackage{float} 
\usepackage[normalem]{ulem} 
\usepackage[shortlabels]{enumitem} 
\usepackage{multirow} 
 
\definecolor{myred}{RGB}{255,50,50} 
\definecolor{myblue}{RGB}{0,0,180}    
\definecolor{mygrey}{RGB}{184,134,11}  
\definecolor{green}{RGB}{0,180,0}

\newcommand{\sgra}	{Sgr~A*} 
\newcommand{\sgrab}	{Sgr~A*~} 

\newcommand*{\RedBF}[1]{\textbf{\textcolor{myred}{#1}}}
 
\def\masy    {mas~yr$^{-1}$}

\def\uas     {$\mu$as}

\def\h       {\ifmmode{^{\rm h}}\else{$^{\rm h}$}\fi} 
\def\m       {\ifmmode{^{\rm m}}\else{$^{\rm m}$}\fi} 
\def\s       {\ifmmode{^{\rm s}}\else{$^{\rm s}$}\fi} 
\def\deg     {\ifmmode{^{\circ}}\else{$^{\circ}$}\fi} 
\def\decdeg  {\ifmmode{{\rlap.}^{\circ}} \else ${\rlap.}^{\circ}$\fi} 
\def\decs    {\ifmmode{{\rlap.}^{\rm s}} \else ${\rlap.}^{\rm s}$\fi} 
\def\decas   {\ifmmode{{\rlap.}{''}}\else{${\rlap.}{''}$}\fi} 
\def\Ho  {\ifmmode{{\rm H}_0}\else{H$_0$}\fi} 
\def\Ro  {\ifmmode{{\rm R}_0}\else{R$_0$}\fi} 
\def\To  {\ifmmode{\Theta_0}\else{$\Theta_0$}\fi} 
 
\def\Vsbar {\ifmmode {\overline{V_s}}\else {$\overline{V_s}$}\fi} 
\def\Usbar {\ifmmode {\overline{U_s}}\else {$\overline{U_s}$}\fi} 
\def\Wsbar {\ifmmode {\overline{W_s}}\else {$\overline{W_s}$}\fi} 
 
\def\mux    {\ifmmode {\mu_x}\else {$\mu_x$}\fi} 
\def\muy    {\ifmmode {\mu_y}\else {$\mu_y$}\fi} 
\def\mura   {\ifmmode {\mu_{\alpha}}\else {$\mu_{\alpha}$}\fi} 
\def\mude   {\ifmmode {\mu_{\delta}}\else {$\mu_{\delta}$}\fi} 
 
\def\gax{\mathrel{\rlap{\lower4pt\hbox{\hskip1pt$\sim$}} 
    \raise1pt\hbox{$>$}}} 
 
\def\d    {\ifmmode {{\rlap{.}}^\circ}\else {${\rlap{.}}^\circ$}\fi} 
\def\s    {\ifmmode {{\rlap{.}}^s}\else {${\rlap{.}}^s$}\fi} 
\def\as   {\ifmmode {{\rlap{.}}^{''}}\else {${\rlap{.}}^{''}$}\fi} 
 
\def\uas {$\mu$as} 
\def\masy{mas~yr$^{-1}$}

\def\lax{\mathrel{\rlap{\lower4pt\hbox{\hskip1pt$\sim$}} 
    \raise1pt\hbox{$<$}}}                
\def\gax{\mathrel{\rlap{\lower4pt\hbox{\hskip1pt$\sim$}} 
    \raise1pt\hbox{$>$}}}                
 
\def\aap{A\&A}

\def\apjs{ApJS} 
\def\apjl{ApJ}

\newcommand{\masyr}	{mas yr$^{-1}$} 
 
\newcommand{\uasyr}	{\ifmmode {\mu{\rm as~yr}^{-1}}\else {$\mu$as~yr$^{-1}$}
\fi}

\shorttitle{Position of \sgra} 
\shortauthors{Xu et al.} 
\graphicspath{{./}{figures/}} 
 
\begin{document} 
\begin{CJK*}{UTF8}{gbsn} 
\title{A Milliarcsecond-accurate Position for Sagittarius A*} 
 
\correspondingauthor{Shuangjing Xu} 
\email{sjxuvlbi@gmail.com} 
\author[0000-0003-2953-6442]{Shuangjing Xu 
} 
\affiliation{Korea Astronomy and Space Science Institute, 776 Daedeok-daero,
Yuseong-gu, Daejeon 34055, Republic of Korea}                           
\affiliation{Shanghai Astronomical Observatory, Chinese Academy of
Sciences, 80 Nandan Road, Shanghai 200030, China}                       
 
\author[0000-0003-1353-9040]{Bo Zhang 
} 
\affiliation{Shanghai Astronomical Observatory, Chinese Academy of
Sciences, 80 Nandan Road, Shanghai 200030, China}                       
%
\author[0000-0001-7223-754X]{Mark J.  Reid} 
\affiliation{Center for Astrophysics~$\vert$~Harvard \& Smithsonian,
60 Garden Street, Cambridge, MA 02138, USA}                             
 
\author{Xingwu Zheng 
} 
\affiliation{ 
School of Astronomy and Space Science, Nanjing University, 22 Hankou
Road, Nanjing  210093, China}                                           
 
\author{Guangli Wang 
} 
\affil{Shanghai Astronomical Observatory, Chinese Academy of Sciences,
80 Nandan Road, Shanghai 200030, China}                                 
 
\author{Taehyun Jung} 
\affiliation{Korea Astronomy and Space Science Institute, 776 Daedeok-daero,
Yuseong-gu, Daejeon 34055, Republic of Korea}                           
 
\begin{abstract} 
The absolute position of Sgr A*, the compact radio source at the center of
the Milky Way, had been uncertain by several tens of milliarcseconds.
Here we report improved astrometric measurements of the absolute
position and proper motion of \sgra.  Three epochs of phase-referencing
observations were conducted with the Very Long Baseline Array for \sgrab at 22 and 43 GHz in
2019 and 2020.  Using extragalactic radio sources with submilliarcsecond-accurate positions as reference, we determined the absolute position of Sgr A*
at a reference epoch 2020.0 to be at
$\alpha$(J2000) = $17^{\rm h} 45^{\rm m}$40\decs032863$~\pm~$0\decs000016
and  $\delta$(J2000) = $-29^\circ 00^\prime 28\decas24260~\pm~$0\decas00047,
with an updated proper motion $-3.152~\pm~0.011$ and $-5.586~\pm~0.006$~\masyr\
in the easterly and northerly directions, respectively.      
\end{abstract} 
 
 
\section{Introduction} \label{sec:intro} 
 
Sagittarius A* (\sgra) is a bright and compact radio source with overwhelming
evidence for being a supermassive black hole (SMBH) at the dynamical center of
the Galaxy
\citep{2008ApJ...689.1044G,2009ApJ...692.1075G,2020ApJ...892...39R,2022ApJ...930L..12E}.
Radio astrometry plays a critical role in the study of \sgra, since it
is precluded from optical view by $\sim30$ mag                   
of visual extinction \citep{2003ApJ...594..294S} and has in the past only been
detected in the infrared (IR) when flaring \citep{2003Natur.425..934G}.     
The absolute positions of compact radio sources are routinely obtained
to an accuracy of milliarcseconds using geodetic and astrometric
Very Long Baseline Interferometry (VLBI) observations at 8.4/2.3 GHz
\citep[e.g.,][]{1998AJ....116..516M}.                                        
However, at these frequencies \sgrab is heavily resolved on
interferometer baselines longer than several
hundreds of km, owing to scatter broadening in the interstellar medium
\citep{1998ApJ...508L..61L}.
Prior to 2004, the absolute position of \sgra, determined with
the Very Large Array (VLA) and a small number of VLBI antennas at 86 GHz
\citep{1992A&A...258..295M,1994ApJ...434L..59R,1999ApJ...518L..33Y}, 
was accurate from about $\pm$200 mas to about $\pm$20 mas.
Next, improved positions for 
two calibrators (J1745--2820 and J1748--2907) within $\sim0.7\deg$ of \sgrab
were obtained from VLBI observations at 8.4/2.3 GHz.  These positions,
also limited by interstellar scattering,
were estimated to be accurate at the $\sim10$ mas level and were used
by \citet{2004ApJ...616..872R,2020ApJ...892...39R} to obtain an improved
position for \sgrab (see coordinates in their Table 1 Notes).

Differential astrometry using VLBI observations involves measuring the
  interferometer phase on a strong ``reference" source and subtracting that phase
  from all sources; this is known as phase referencing \citep{1995ASPC...82..327B}.
  This removes interferometer coherence-time limitations and allows weak sources
  to be imaged and relative positions measured with high accuracy
  \citep{2014ARA&A..52..339R}.
An inaccurate absolute position of the source used as the interferometer phase
reference in VLBI observations 
can cause an error in its delay/phase
measurements, and these errors will be propagated to the target source
\citep{2014ARA&A..52..339R,2017isra.book.....T}.
If the reference source position error is large enough ($\gax10$
mas), there can be significant second-order effects which lead to
a relative position error and degraded image quality \citep{2014ARA&A..52..339R}.
These errors become more serious for mm/sub-mm VLBI observations
\citep{2022ApJ...930L..12E} and phase-referenced infrared interferometry
observations \citep{2017A&A...602A..94G} of \sgra,                      
since a delay error leads to an interferometer phase error scaled by the
observing frequency.

Recently the spatial resolution and sensitivity of IR and radio telescopes
has significantly improved.  The GRAVITY instrument on the Very Large
Telescope Interferometer (VLTI) has achieved mas-scale resolution
\citep{2017A&A...602A..94G}, and the Atacama Large Millimeter/Submillimeter
Array (ALMA) can also perform precision astrometry of the Galactic center
region around \sgrab with high angular resolution over wide
field of view \citep{2022arXiv220406778T}.                              
Therefore, an accurate absolute position of \sgrab will be valuable
for the association of sources detected at different times or wavelengths.

In this paper, we report VLBA phase-referenced observations
designed to improve the accuracy of the absolute position of \sgra.  
Compared to \citet{2004ApJ...616..872R,2020ApJ...892...39R},
we used more distant ($\sim2\deg$ separation) calibrators as position references
in order to minimize the effects of scatter broadening.  Using these
calibrators we arrive at a mas-accurate position for \sgra.

\section{Observations and Data Reduction} 
\label{obs} 
 
\begin{deluxetable*}{cllrrrlll} 
\tablenum{1} 
\tablecaption{VLBA Astrometric  Observations of \sgrab in 2019 and
2020  \label{tab:obs}}                                                  
\tablewidth{0pt} 
\tablehead{ 
\colhead{Epoch} & \colhead{Program}  & \colhead{Stations\tablenotemark{a}}
&  \colhead{Duration} & \colhead{}& \colhead{Frequency\tablenotemark{b}}
& \\                                                                    
\colhead{(yyyy-mm-dd)} & \colhead{Code} & \colhead{} & \colhead{(hour)}
& \colhead{} & \colhead{Band}  &                                        
} 
\startdata 
2019-06-18 & BX008A & ALL except  SC & 7 &  & K/Q &  \\ 
2019-11-01 & BX008B & ALL & 7 &  & K/Q  &  \\ 
2020-01-26 & BX008C & ALL except  MK and SC & 7 &  & Q &\\ 
\enddata 
\tablenotetext{a}{VLBA has 10 stations located at Saint Croix (SC), 
Hancock (HN), North Liberty (NL), Fort Davis (FD), Los Alamos (LA), 
Pie Town (PT), Kitt Peak (KP), Owens Valley (OV), Brewster (BR), 
Mauna Kea (MK).} 
\tablenotetext{b}{The frequency band K is 22 GHz and Q is 43 GHz. 
Only Q band was used for \sgrab astrometry in BX008C.} 
\end{deluxetable*} 
 
\begin{deluxetable}{llllrrl} 
\tablenum{2} 
  \tablecolumns{8} 
  \tablewidth{0pc} 
  \tablecaption{Initial Source Positions Used in the VLBA Program BX008} 
  \tablehead{ 
  \colhead{Source} &  \nocolhead{ID\tablenotemark{b} } & \colhead{R.A.
(J2000)} & \colhead{Dec.  (J2000)}                   & \colhead{$\theta_{sep}$}
& \colhead{P.A.}     & \colhead{Position Reference}   \\                
    \colhead{      } &  \colhead{      } & \colhead{h~~~m~~~s } 
& \colhead{\degr~~~\arcmin~~~\arcsec\ }  & \colhead{(\degr)}    
& \colhead{(\degr)}  & \colhead{}                                       
  } 
  \startdata 
  \sgra \tablenotemark{a}   &   & 17:45:40.0353\tablenotemark{a}
  & $-$29:00:28.247\tablenotemark{a}    & ... & ... &
  \citet{2004ApJ...616..872R,2020ApJ...892...39R} \\                                                                      
  J1745-2820  &  & 17:45:52.4968   &  $-$28:20:26.294
&  0.7 & 4    &  \citet{2004ApJ...616..872R,2020ApJ...892...39R}  \\    
  J1748-2907  &  & 17:48:45.6860    &  $-$29:07:39.404
&  0.7 & 100 &  \citet{2004ApJ...616..872R,2020ApJ...892...39R}  \\     
  J1752-2956  &  & 17:52:33.1081   &  $-$29:56:44.916
&  1.8 & 122 & rfc$\_$2016c\tablenotemark{b}  \\ 
  J1744-3116  &  & 17:44:23.5782   &  $-$31:16:36.294
&  2.3 & 187  & rfc$\_$2016c\tablenotemark{b}   \\    
  \enddata 
  \tablecomments{ 
  $\theta_{sep}$ and P.A. indicate source separations and position 
  angles (East of North) from \sgra. 
  \tablenotetext{a}{The expected position of \sgrab at the epoch
2019-06-18, which was calculated using the position (17:45:40.04091,
$-$29:00:28.1175) on 1996-03-20 and an assumed proper motion  $-3.147$
and $-5.578$ \masyr\ in the easterly and northerly directions, respectively.}
\tablenotetext{b}{
These positions were from the Radio Fundamental Catalog (RFC) version of rfc\_2016c (\url{http://astrogeo.org/vlbi/solutions/rfc_2016c/}).
Both J1752--2956 and J1744--3116 had positional accuracy
of approximately $\pm1$ mas level. 
Source coordinates listed in this Table were rounded to the nearest mas and
adopted during the observations/correlations; these were updated later in data
analysis as listed in Table~\ref{tab:new_pos}.
} } 
\label{tab:aips_pos} 
\end{deluxetable} 
 
\begin{figure*}[t] 
\epsscale{0.7} 
\plotone{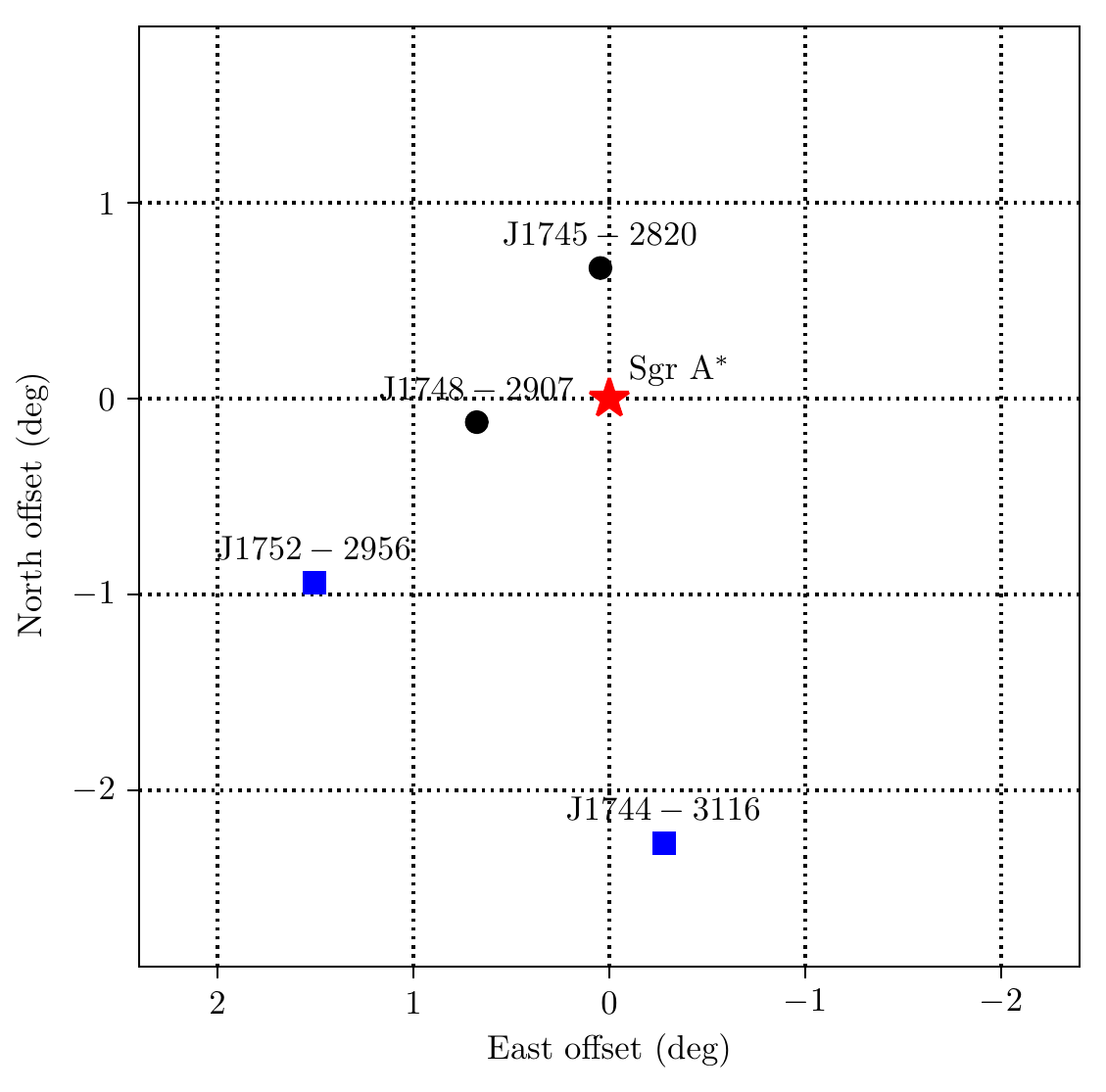} 
\caption{The spatial distribution of sources from the VLBA program BX008. 
\sgrab({\it red star}) is the interferometer phase reference; two nearby sources
({\it black dots}) are from \citet{2004ApJ...616..872R,2020ApJ...892...39R};
and two more distant calibrators ({\it blue squares}) are from the
rfc\_2016c catalog.   
J1744-3116 and J1752-2956 have accurate absolute positions  ($\sim1$ mas accuracy in rfc\_2016c catalog,  sub-mas accuracy in rfc\_2022a catalog)
and can be used to improve the position of \sgra.
The other two calibrators J1745--2820 and J1748--2907 are used to improve
the proper motion of \sgra\ as done in \citet{2020ApJ...892...39R}.
\label{fig:sou}} 
\end{figure*} 
 
We conducted three epochs of phase-referenced observations of \sgrab and four quasars
at 22 (K-band) and 43 (Q-band) GHz under the National Radio Astronomy Observatory
\footnote{The National Radio Astronomy Observatory is operated by
Associated Universities Inc., under a cooperative agreement with the
National Science Foundation.}  program BX008 as listed in Table \ref{tab:obs}.  
Four extragalactic radio sources shown in Figure~\ref{fig:sou}
and listed in Table~\ref{tab:aips_pos} were used                                 
as background references for astrometry.
The calibrators, J1744-3116 and J1752-2956, separated by $\approx2$\deg\
have $\sim1$ mas accurate absolute positions in the rfc\_2016c\footnote{The Radio Fundamental
Catalog (RFC) with the version of rfc\_2016c
(\url{http://astrogeo.org/vlbi/solutions/rfc_2016c/})} catalog and were used to improve the
absoltue position of \sgra.
Note, the absolute positions of these calibrators used during the observations
and in the VLBI correlator were updated later in data processing as more accurate (sub-mas accuracy) 
positions 
became available in the rfc\_2022a\footnote{The Radio Fundamental
Catalog (RFC) with the version of rfc\_2022a
(\url{http://astrogeo.org/sol/rfc/rfc_2022a/})} catalog (see Section \ref{sec_pos}).
The nearby calibrators, J1745--2820 and J1748--2907, separated from \sgrab\ by
$0.7$\deg, were used to improve the proper motion of \sgrab by combining with the
data presented in \citet{2020ApJ...892...39R}.  

We observed at K and Q band alternatively in each session (except for
BX008C where we only observed at Q band) and  
switched between \sgrab and a calibrator source every 40/20 seconds 
for K/Q band.  We used \sgrab as the phase-reference, because
it is stronger than the background sources and could be detected
on individual baselines within a scan. 
Due to interstellar scattering, \sgrab was heavily resolved on the 
longer baselines associated with antennas at Saint Croix (SC), Hancock
(HN), and Mauna Kea (MK).  The Kitt Peak (KP) antenna in program BX008C
failed for unknown reasons. 
For each session we used 7-hour tracks.  We included four $\sim$25-min
``geodetic blocks'' at K-band scheduled before, middle and after the phase-reference
observations, which allowed us to calibrate tropospheric and clock delays. 
Three strong ``fringe finders,'' J1638+5720, J1733-1304 (NRAO530) and J2236+2828,
were also observed to monitor delay and electronic phase differences 
among the intermediate-frequency (IF) bands.   We observed with four 64-MHz IF bands and 
recorded both right and left circularly polarized signals with Nyquist 
sampling and 2 bits-per-sample for a total sampling rate of 2 Gbps. 
 
The raw data were processed with the DiFX correlator \citep{2007PASP..119..318D}
in Socorro, NM, which generated 128 spectral channels for
each 64-MHz IF band.  Data calibration was performed with the NRAO Astronomical Image
Processing System (AIPS) \citep{2003ASSL..285..109G}.                   
Amplitude calibration was done using system noise temperatures and
antenna gains logged during observations.                               
For phase calibration, we corrected for the effects of diurnal feed
rotation and errors in the Earth Orientation Parameters (EOP).          
Ionospheric delays were corrected for with the Global Ionosphere Map provided by the International GNSS Service \citep{2014ARA&A..52..339R}. 
After calibration, we imaged all sources and measured their position
offsets relative to \sgrab by fitting elliptical Gaussian brightness
distributions. 
 
\section{Results} \label{results}

\begin{deluxetable}{ccrrr|rrrr|rrr} 
\tablenum{3} 
\tablewidth{0pt} 
\tabletypesize{\small} 
\tablecaption{Residual Position Offsets Relative to \sgrab in Program
BX008}                                                                  
\tablehead{ 
\colhead{Source} & \colhead{Epoch}   & \colhead{Frequency} & 
\colhead{East Offset A\tablenotemark{a} } & \colhead{North Offset
A\tablenotemark{a}}&                                                    
\colhead{East Offset B\tablenotemark{b}}& \colhead{North Offset B\tablenotemark{b}}
\\                                                                      
\colhead{}               & \colhead{}   & \colhead{Band} & 
 \colhead{(mas)} & \colhead{(mas)}    &\colhead{(mas)}       & \colhead{(mas)}
 } 
\startdata 
J1745--2820 & 2019.461  & Q  &$-0.11\pm0.06$ &$ 0.34\pm0.13$ &$-0.11\pm0.05$ & $ 0.42\pm0.12$ \\
                       && K  &$-0.13\pm0.07$ &$ 0.25\pm0.14$ &$-0.07\pm0.06$ & $ 0.34\pm0.12$ \\
            & 2019.835  & Q  &$ 1.34\pm0.03$ &$ 2.58\pm0.06$ &$ 1.36\pm0.03$ & $ 2.56\pm0.06$ \\
                       && K  &$ 1.26\pm0.05$ &$ 2.44\pm0.08$ &$ 1.26\pm0.05$ & $ 2.43\pm0.08$ \\
            & 2020.070  & Q  &$ 1.86\pm0.06$ &$ 4.10\pm0.12$ &$ 1.86\pm0.06$ & $ 4.21\pm0.14$ \\
\hline 
J1748--2907 & 2019.461  & Q  &$ 1.45\pm0.05$ &$-1.52\pm0.12$ &$ 1.43\pm0.04$ & $-1.22\pm0.11$ \\
                       && K  &$ 1.58\pm0.05$ &$-1.75\pm0.09$ &$ 1.58\pm0.04$ & $-1.44\pm0.09$ \\
            & 2019.835  & Q  &$ 2.85\pm0.04$ &$ 0.35\pm0.10$ &$ 2.78\pm0.02$ & $ 0.76\pm0.07$ \\
            & 2020.070  & Q  &$ 2.98\pm0.08$ &$ 1.41\pm0.21$ &$ 3.03\pm0.06$ & $ 1.75\pm0.20$ \\
\hline 
J1752--2956 & 2019.835  & K &$31.36\pm0.10$ &$-5.32\pm0.26$  &$31.29\pm0.03$ & $-4.63\pm0.11$ \\
\hline 
J1744--3116 & 2019.461  & K &$30.19\pm0.10$ &$-5.73\pm0.25$  &$30.18\pm0.09$ & $-5.79\pm0.23$ \\
            & 2019.835  & Q &$31.48\pm0.07$ &$-2.48\pm0.17$  &$31.47\pm0.06$ & $-2.49\pm0.18$ \\
                       && K &$31.36\pm0.09$ &$-2.27\pm0.20$  &$31.39\pm0.08$ & $-2.30\pm0.19$ \\
\enddata 
\tablenotetext{a}{Case A: Position offsets relative to \sgra\ in the correlated
interferometer visibilities.  The coordinates used in correlation 
are in Table \ref{tab:aips_pos}. The formal position
uncertainties reflect thermal (random) noise in the phase-referenced
interferometric images but do not include systematic uncertainties.}    
\tablenotetext{b}{Case B: same as Case A, but correcting for
second-order effects (see Section \ref{results})
of the position offset ([$-$31,+2] mas) we find for \sgra. 
}                                                            
  \label{tab:measured_pos} 
\end{deluxetable}

\begin{figure*}[t] 
\epsscale{0.8} 
\plotone{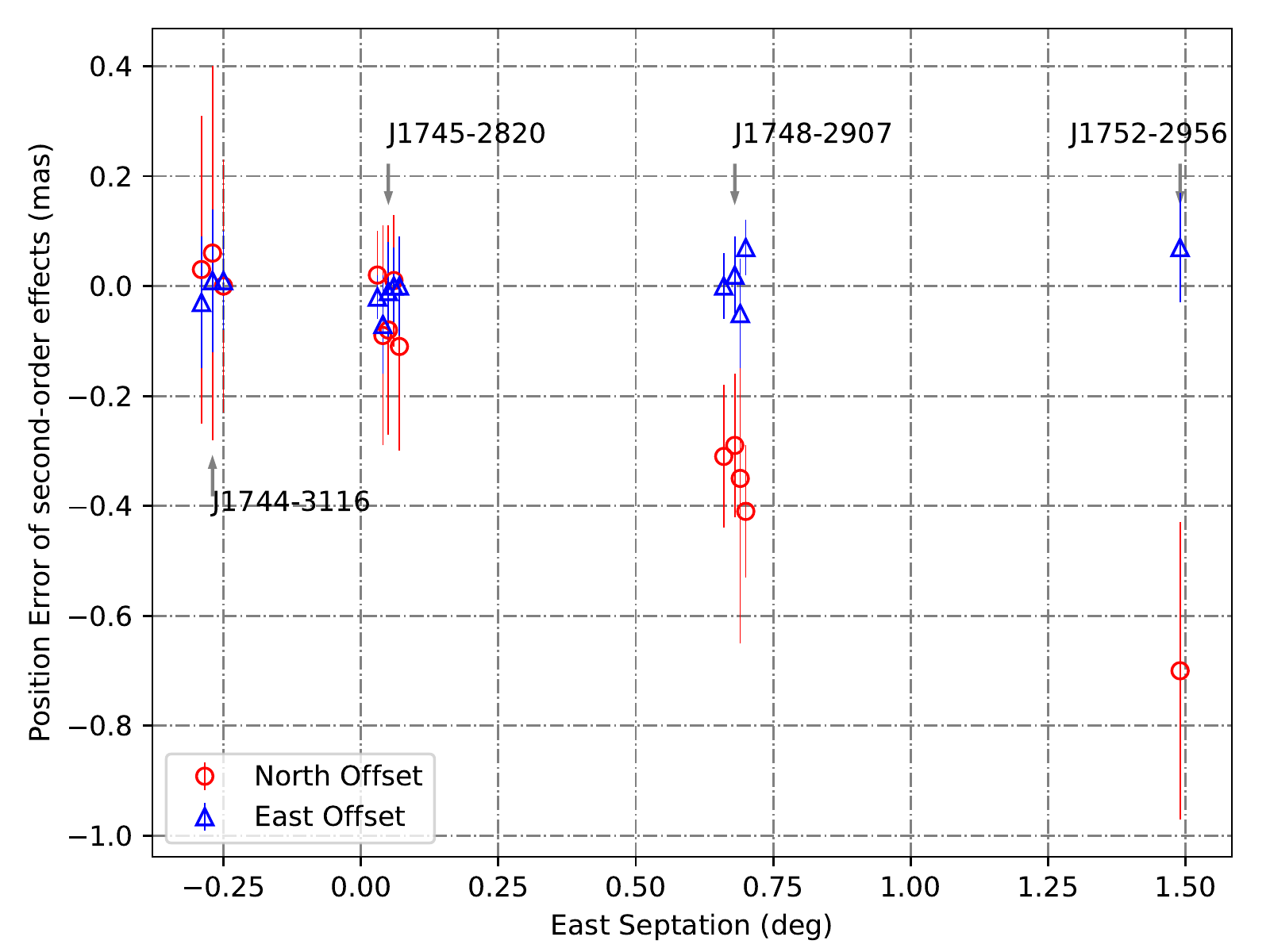} 
\caption{The position error due to the second-order effects of the
position offset ([$-$31,+2] mas) of \sgra. The error bars are 
joint thermal uncertainties from phase-referenced images
with and without the position offset correction.                                        
The horizontal axis is the separation angles in eastern direction
between each calibrator and \sgra.  Data from
different epochs and different bands are offset slightly horizontally
for clarity.                                                            
\label{fig:second-order}} 
\end{figure*} 
 
Table \ref{tab:measured_pos} lists the measured position offsets relative to \sgra. 
Notice that the position offsets of the nearby (in angle)
calibrators J1745--2820 and J1748--2907 are small.  This
is an historical artifact from previous observations which used 
inaccurate positions for the calibrators to determine the position of \sgra.
The calibrators J1752--2956 and J1744--3116 display large position offsets
of $\approx30$ mas {\it relative to \sgra}. 
As mentioned earlier, if the interferometer phase-reference source
has a position error, this will be transferred to the other sources.
Since J1752--2956 and J1744--3116 have very accurate absolute positions ($\sim1$ mas in Table \ref{tab:aips_pos})
determined independently of \sgra, we conclude that the absolute positions of \sgra,
J1745--2820, and J1748--2907 used by
\citet{2004ApJ...616..872R,2020ApJ...892...39R} are in error by
the reverse of the offsets seen for J1752--2956 and J1744--3116. 

The problem with obtaining an accurate absolute position for Sgr A* is the
  extremely large scatter broadening experienced by sources within about $1\deg$
  of the Galactic center. This makes previous X/S-band (8.4/2.3 GHz) VLBI astrometry,
  which provides absolute positions for a large number of calibrators, only marginally
  productive, since only very short baselines could be used to produce
  multi-band delays in that part of the sky.  It took our study, which bootstrapped
  positions from calibrators outside of the large scattering region to obtain a
  mas-accurate position for Sgr A* and the calibrators used by
  \citet{2004ApJ...616..872R,2020ApJ...892...39R}.

The position offsets or errors in the source used as the phase reference can cause errors in its delay/phase measurements, which are then propagated to the target source. 
For the first-order effects, the offset in the reference source position is simply transferred to the estimate of the target source position \citep{2017isra.book.....T}.
However, if the reference source position offset becomes large ($\gax10$ mas), additional second-order effects can be important
 due to the fact that the target and reference sources are separated on the sky. Because they are at different positions, the interferometer phase response to a position offset at one position on the sky does not exactly mimic the response at the other position, which can lead to a small position error and to degraded image quality \citep{2014ARA&A..52..339R}.
In order to evaluate these corrections, we re-processed the data using a
corrected position of \sgra, shifted by ($-$31,+2) mas using the AIPS task ``CLCOR,''
and re-measured the relative positions.   We then applied the same amplitude
and phase calibrations described in Section \ref{obs}.    
Figure \ref{fig:second-order} shows the second-order position errors, which were
obtained by subtracting the original from the re-measured positions.
These corrections were as large as $-0.7$ mas, and the last two columns in
Table \ref{tab:measured_pos} have these applied.

\subsection{The Absolute Position of \sgra} \label{sec_pos} 

In active galactic nuclei (AGN) jets, the core of the radio emission can
  shift as a function of observing frequency owing to differences in the opacity
  \citep{1979ApJ...232...34B, 2011Natur.477..185H}; this is known as the
  ``core-shift'' effect.
The differences in position offsets measured at K and Q band at the
same epoch generally are $\lax0.2$ mas and are consistent within their
joint uncertainties (see Table \ref{tab:measured_pos}). 
This is consistent with
the lack of a frequency-dependent core-shift between K and Q band for \sgrab
(which would lead to an {\it apparent} core-shift in the calibrators)
reported by \citet{2015ApJ...798..120B} at a comparable level.
Therefore, we average the K and Q band results when estimating
absolute positions.

In order to estimate the absolute position of \sgra, we do the following:
First, we corrected the data for improved positions for J1744--3116 and J1752--2956,
compared to those used during the correlation of the VLBA data, using
the latest RFC version 
rfc\_2022a
(see Table \ref{tab:new_pos}).
Second, we correct the individual position offsets to a reference epoch 2020.0,
assuming \sgrab has an easterly motion $-3.156\pm0.006$~\masyr\ and a northerly
motion $-5.585\pm0.010$~\masyr\ \citep{2020ApJ...892...39R}; we also correct            
for \sgra's parallax of 0.123 mas assuming a distance of $8.15$ kpc
\citep{2019A&A...625L..10G,2019ApJ...885..131R}.
Third, we correct for second-order effects of correlating the data with the
``old'' \sgrab position, which is in error by about 30 mas (i.e., ``Case
B'' in Table \ref{tab:measured_pos}).  
Fourth, we average the position offsets 
at the reference epoch 2020.0, and infer a shift for \sgrab of
($-32.26\pm0.24$, $4.28\pm0.53$) mas relative to J1752--2956 and
($-31.29\pm0.37$, $4.83\pm1.02$) mas relative to J1744--3116.\footnote{The quoted uncertainties include the following added in quadrature:
i) thermal noise of ($\pm0.03,\pm0.11$) mas for J1752--2956 and
($\pm0.08,\pm0.20$) mas for J1744--3116; ii) the propagated error from proper motion
and parallax ($\pm0.01,\pm0.01$) mas of \sgra;  iii) the uncertainties in the
absolute positions of ($\pm0.20,\pm0.40$) mas for J1752--2956 and
($\pm0.36,\pm0.70$) mas for J1744--3116\RedBF{)}, and iv) estimated uncertainties
from phase-referencing ($\pm0.12,\pm0.33$) mas for J1752--2956
and ($\pm0.06,\pm0.71$) mas for J1744--3116 based on \citet{2006A&A...452.1099P}.
}
Finally, we average (variance weighting) these position offsets
 and reverse their signs to arrive at our best position of \sgra.
Similarly, relative to the best positions of \sgra, we average the position offsets of the other two calibrators  
at the reference epoch 2020.0, and infer a shift for J1745--2820 of ($-30.28\pm0.22$, $7.93\pm0.53$) mas and for J1748--2907 of 
($-28.84\pm0.23$, $6.00\pm0.49$) mas\footnote{The quoted uncertainties include the following added in quadrature:
i) thermal noise of ($\pm0.05,\pm0.10$) mas for J1745--2820 and
($\pm0.04,\pm0.12$) mas for J1748--2907; ii)  the uncertainties in the
absolute positions of ($\pm0.21,\pm0.47$) mas for the updated \sgra, and iii) estimated uncertainties
from phase-referencing ($\pm0.03,\pm0.23$) mas for J1745--2820
and ($\pm0.08,\pm0.06$) mas for J1748--2907 based on \citet{2006A&A...452.1099P}.
}.
Our best positions for \sgrab and the four calibrators are given in Table
\ref{tab:new_pos}.

Recently, using geodetic and astrometric VLBA observations at 24 GHz,
\cite{2022ivs..meetE...1G} \footnote{\url{https://drive.google.com/file/d/1350Itc6wVb14HzrcoRzn-Ew-L-x_VwL3/view}}
reported an absolute position of \sgrab at a reference epoch 2015.0 to be at
$\alpha$(J2000) = $17^{\rm h} 45^{\rm m}$40\decs034051$~\pm~$0\decs000019
and  $\delta$(J2000) = $-29^\circ 00^\prime 28\decas21583~\pm~$0\decas00045,
and a proper motion $-3.144~\pm~0.044$ and $-5.626~\pm~0.080$~\masyr\
in the easterly and northerly directions, respectively.   
Transforming the reference epoch from 2015.0 to 2020.0, results in its position of
$\alpha$(J2000) = $17^{\rm h}45^{\rm m}$40\decs032853$\pm$0\decs000025
and
$\delta$(J2000) = $-29^\circ 00^\prime 28\decas24396\pm$0\decas00060, which is consistent with our results  within a 1-$\sigma$ error in $\alpha$ and a 2-$\sigma$ error in $\delta$, respectively.

\begin{deluxetable}{llllrll} 
\tablenum{4} 
  \tablecolumns{8} 
  \tablewidth{0pc} 
  \tablecaption{The Updated Sources Positions} 
  \tablehead{ 
  \colhead{Source}  & \colhead{R.A. (J2000)} & \colhead{Dec.  (J2000)} 
   & \colhead{Epoch}     &  \colhead{Reference}   \\ 
    \colhead{      }  & \colhead{h~~~m~~~s ($\pm$ s)}  & \colhead{\degr~~~\arcmin~~~\arcsec\
($\pm$ \arcsec)}   & \colhead{}  & \colhead{}  & \colhead{}             
  } 
  \startdata 
J1752--2956   & 17:52:33.108059 ($\pm$0.000013)   &  $-$29:56:44.91541 ($\pm$0.00040)  & 2000.0 & rfc$\_$2022a\tablenotemark{a} \\       
J1744--3116   & 17:44:23.578245 ($\pm$0.000024)   &  $-$31:16:36.29107 ($\pm$0.00070)  & 2000.0  & rfc$\_$2022a\tablenotemark{a} \\       
\hline 
\sgra  & 17:45:40.032863 ($\pm$0.000016) & $-$29:00:28.24260 ($\pm$0.00047)
& 2020.0    & This Paper \\ 
J1745--2820  & 17:45:52.494492 ($\pm$0.000017)    &  $-$28:20:26.28607 ($\pm$0.00053) & 2020.0  &  This Paper \\                
J1748--2907  & 17:48:45.683802 ($\pm$0.000018)    &  $-$29:07:39.39800 ($\pm$0.00049) &2020.0  &  This Paper \\
\enddata 
 \tablenotetext{a} 
 {The updated positions of J1752--2956 and J1744--3116 are from
the Radio Fundamental Catalog (RFC) version rfc\_2022a
(\url{http://astrogeo.org/sol/rfc/rfc_2022a/}), which
had better accuracy than the positions in Table \ref{tab:aips_pos}
from version of rfc\_2016c.}
  \label{tab:new_pos} 
\end{deluxetable} 

\subsection{The Updated Proper Motion of \sgra} \label{sec_plx}

Our Q band data for J1745--2820 and J1748--2907 can also be used
to update the proper motion of \sgrab of \citet{2020ApJ...892...39R}
by extending the time spanned to 25 years (1995--2020).
The prior measurements are equivalent to
``Case A'' in Table \ref{tab:measured_pos}, and in order to be
internally consistent, we list the positions of the two
quasars relative to \sgrab in Table \ref{tab:RB2020_pos}.
Note that these do {\it not} use our updated positions for \sgrab and
J1745--2820 and J1748--2907.
The position uncertainties include estimates of systematic effects,
dominated by small residual errors in modeling atmospheric delays.      
 
\begin{deluxetable}{ccrrrrrr} 
\tablenum{5} 
\tablewidth{0pt} 
\tabletypesize{\small} 
\tablecaption{Residual Position Offsets Relative to \sgrab for updating
the                                                                     
proper motion of \citet{2020ApJ...892...39R}} 
\tablehead{ 
\colhead{Source}         & \colhead{Date of}      & \colhead{Band} & 
\colhead{East Offset}& \colhead{North Offset} & 
 \\ 
\colhead{}               & \colhead{Observation}   & \colhead{} & 
\colhead{(mas)}       & \colhead{(mas)    }  & 
 } 
\startdata 
J1745--2820 & 2019.461  & Q &$ 73.35\pm 0.1$ &$ 129.34\pm 0.2$ & \\ 
            & 2019.835  & Q &$ 74.80\pm 0.1$ &$ 131.58\pm 0.2$ & \\ 
            & 2020.070  & Q &$ 75.32\pm 0.1$ &$ 133.10\pm 0.2$ & \\ 
J1748--2907 & 2019.461  & Q &$ 74.91\pm 0.2$ &$ 127.48\pm 0.6$ & \\ 
            & 2019.835  & Q &$ 76.31\pm 0.2$ &$ 129.35\pm 0.6$ & \\ 
            & 2020.070  & Q &$ 76.44\pm 0.2$ &$ 130.41\pm 0.6$ & \\ 
\enddata 
\tablecomments{ 
  The coordinate offsets are relative to the following J2000 positions
  used by \citet{2020ApJ...892...39R}: 
\sgrab (17 45 40.0409, --29 00 28.118), 
J1745--283 (17 45 52.4968, --28 20 26.294), and 
J1748--291 (17 48 45.6860, --29 07 39.404).
  \label{tab:RB2020_pos} 
} 
\end{deluxetable} 
 
Before fitting for motions, we made a small correction for the parallax effect ($<=0.123$ mas),
assuming a distance to the Galactic center of $8.15$ kpc
\citep{2019A&A...625L..10G,2019ApJ...885..131R}, to all the measurements.
Then fitting apparent motions of \sgrab relative to J1745--2820 and J1748--2907 using variance-weighted
least-squares we find the motions listed in Table \ref{table:motions}.
The northerly motions measured against the two quasars are statistically
consistent.  However, in the easterly direction the motions formally differ by $0.021\pm 0.0064$ \masy,
suggesting that there might be a small sytematic difference between them.   One possibility is
one or both quasars are not stationary point sources, owing to jet emissions within our resolution
element.   In order to allow for systematics, we add one-half the difference in the motion
measurements relative to the two quasars in quadrature with the formal uncertainties.
This yields a combined motion estimate for \sgrab of $-3.152\pm0.011$ \masy\ in the easterly
direction and $-5.586\pm0.006$ \masy\ in the northerly direction.
 
\begin{deluxetable}{rcccc} 
\tablewidth{0pt} 
\tablenum{6} 
\tabletypesize{\small} 
\tablecaption{The Updated Apparent Relative Motions} 
\tablehead{ 
\colhead{Source -- Reference}         & 
\colhead{Easterly Motion}   & \colhead{Northerly Motion} \\ 
\colhead{}               & 
\colhead{(mas y$^{-1}$)}       & \colhead{(mas y$^{-1}$)}  
 } 
\startdata 
\sgra~--~J1745--2820 (1995--2020)&$-3.141\pm0.005$  &$-5.585\pm0.007$ \\         
\sgra~--~J1748--2907 (1995--2020)&$-3.162\pm0.004$  &$-5.588\pm0.012$ \\    
\hline
\sgra~--~Combined    (1995--2020)&$-3.152\pm0.011$  &$-5.586\pm0.006$ \\         
\enddata 
\tablecomments{Motions values are from weighted least-squares fits
to the data in Table 1 of \citet{2020ApJ...892...39R} and Table \ref{tab:RB2020_pos},
with uncertainties scaled  to give a reduced chi-squared of unity.   
The scale factors for the uncertainties are 0.92 for ``\sgra~--~J1745--2820'' and 0.75
for ``\sgra~--~J1748--2907''.
``Combined'' motions are variance-weighted averages of the individual results
with formal uncertainties (see text for a discussion of systematic uncertainty).
} \label{table:motions} 
\end{deluxetable} 

\section{Conclusions and Outlook} 

We have improved the absolute position and proper motion of \sgrab using
the reference sources with accurate positions.  These calibrators are
separated from \sgrab by $\approx2\deg$ and are much less scatter broadened
than the more nearby sources J1745--2820 and J1748--2907 ($\approx0.7\deg$ from \sgra)
previously used to determine \sgra's position \citep{2004ApJ...616..872R,2020ApJ...892...39R}.
Correcting for a local reference frame offset about 30 mas, we find the
position of \sgrab at the reference epoch 2020.0 to be at
$\alpha$(J2000) = $17^{\rm h} 45^{\rm m}$40\decs032863$\pm$0\decs000016
and
$\delta$(J2000) = $-29^\circ 00^\prime 28\decas24260\pm$0\decas00047.
Transforming the reference epoch from 2020.0 to 2000.0 using \sgra's 
proper motion, results in a position of
$\alpha$(J2000) = $17^{\rm h}45^{\rm m}$40\decs037669$\pm$0\decs000023
and
$\delta$(J2000) = $-29^\circ 00^\prime 28\decas13088\pm$0\decas00049.

By extending the time series of measurements by \citet{2020ApJ...892...39R}
of \sgrab relative to J1745--2820 and J1748--2907, we improved the accuracy
of \sgra's proper motion yielding $-3.152\pm0.011$ and $-5.586\pm0.006$~\masyr\ in the
easterly and northerly directions.         

Our improved absolute position for \sgrab should be used in 
high angular resolution interferometric observations, including VLBI,
VLTI and ALMA observations of the Galactic center region.  If \sgrab is
used as the interferometer phase reference, this will allow higher
dynamic-range imaging and a more accurate registration of images
between radio, millimeter, and infrared observations.
In addition, as shown in Figure \ref{fig:second-order},
the errors in relative position measurements due to the second-order
effect of the inaccurate absolution position can lead small extra position shifts,
which are dependent on the sky positions of target and calibrator.
This can degrade relative astrometry accuracy.  Thus, an
accurate absolute position for
\sgrab will be important for \uas\ relative astrometry in the Galactic center
region, e.g., measuring the parallax of \sgrab with \uas\ accuracy using the
next generation instruments \citep{2020A&ARv..28....6R}.

In addition, the astrometric reference frame for the stars seen
orbiting an unseen mass in the Galactic center at IR wavelengths
is established using common stellar sources seen in radio and infrared images
\citep{1997ApJ...475L.111M,2003ApJ...587..208R,2008ApJ...689.1044G,2009ApJ...692.1075G,2019ApJ...873...65S}. Our improved absolute position for \sgrab provides
an accurate origin for the astrometric reference frame in the Galactic centre region.
For example, using IR measurement of \sgrab flares, this would allow transfer of
the absolute radio to the IR position of \sgrab directly at the mas level. 
  
Finally, we note that the IAU definition of Galactic coordinates
\citep{1960MNRAS.121..123B} places the Galactic center at 
$\alpha_0(B1950)=17^{\rm h}42^{\rm m}26\decs603, \delta_0(B1950)=-28^\circ55^\prime00\decas445$
\citep{1979PASP...91..405L} in B1950.0 coordinates.
Converting the B1950.0 coordinates to J2000.0 coordinates,
\citet{2004ApJ...616..872R} obtained a J2000 origin 
$\alpha_0(J2000)=17^{\rm h}45^{\rm m}37\decs224, \delta_0(J2000)=-28^\circ56^\prime10\decas23$.   However, the IAU definition of Galactic coordinates was based on early HI
observations, prior to the discovery of \sgra.  Given the strong evidence that
\sgrab is a supermassive black hole at the dynamical center of the Galaxy,
one could consider a revised definition which places \sgrab at the origin of
Galactic coordinates.   Note, however, that owing to the Sun's Galactic orbit,
the apparent position of \sgrab drifts by over 6 \masy, which would require a
time-dependent coordinate system.

\acknowledgments 
We are grateful to all staff members in VLBA who helped to operate the array and to correlate the data.
BZ was supported by the National Natural Science Foundation of China (Grant No. U2031212 and U1831136),
and Shanghai Astronomical Observatory, Chinese Academy of Sciences (Grant No. N2020-06-19-005). 
 
\vspace{5mm} 
\facilities{VLBA} 
\software{AIPS \citep{2003ASSL..285..109G}, Astropy \citep{2013A&A...558A..33A}
}

\bibliography{ref}{} 

\begin{thebibliography}{}
\expandafter\ifx\csname natexlab\endcsname\relax\def\natexlab#1{#1}\fi
\providecommand{\url}[1]{\href{#1}{#1}}
\providecommand{\dodoi}[1]{doi:~\href{http://doi.org/#1}{\nolinkurl{#1}}}
\providecommand{\doeprint}[1]{\href{http://ascl.net/#1}{\nolinkurl{http://ascl.net/#1}}}
\providecommand{\doarXiv}[1]{\href{https://arxiv.org/abs/#1}{\nolinkurl{https://arxiv.org/abs/#1}}}

\bibitem[{{Astropy Collaboration} {et~al.}(2013){Astropy Collaboration},
  {Robitaille}, {Tollerud}, {Greenfield}, {Droettboom}, {Bray}, {Aldcroft},
  {Davis}, {Ginsburg}, {Price-Whelan}, {Kerzendorf}, {Conley}, {Crighton},
  {Barbary}, {Muna}, {Ferguson}, {Grollier}, {Parikh}, {Nair}, {Unther},
  {Deil}, {Woillez}, {Conseil}, {Kramer}, {Turner}, {Singer}, {Fox}, {Weaver},
  {Zabalza}, {Edwards}, {Azalee Bostroem}, {Burke}, {Casey}, {Crawford},
  {Dencheva}, {Ely}, {Jenness}, {Labrie}, {Lim}, {Pierfederici}, {Pontzen},
  {Ptak}, {Refsdal}, {Servillat}, \& {Streicher}}]{2013A&A...558A..33A}
{Astropy Collaboration}, {Robitaille}, T.~P., {Tollerud}, E.~J., {et~al.} 2013,
  \aap, 558, A33, \dodoi{10.1051/0004-6361/201322068}

\bibitem[{{Beasley} \& {Conway}(1995)}]{1995ASPC...82..327B}
{Beasley}, A.~J., \& {Conway}, J.~E. 1995, in Astronomical Society of the
  Pacific Conference Series, Vol.~82, Very Long Baseline Interferometry and the
  VLBA, ed. J.~A. {Zensus}, P.~J. {Diamond}, \& P.~J. {Napier}, 327

\bibitem[{{Blaauw} {et~al.}(1960){Blaauw}, {Gum}, {Pawsey}, \&
  {Westerhout}}]{1960MNRAS.121..123B}
{Blaauw}, A., {Gum}, C.~S., {Pawsey}, J.~L., \& {Westerhout}, G. 1960, \mnras,
  121, 123, \dodoi{10.1093/mnras/121.2.123}

\bibitem[{{Blandford} \& {K{\"o}nigl}(1979)}]{1979ApJ...232...34B}
{Blandford}, R.~D., \& {K{\"o}nigl}, A. 1979, \apj, 232, 34,
  \dodoi{10.1086/157262}

\bibitem[{{Bower} {et~al.}(2015){Bower}, {Deller}, {Demorest}, {Brunthaler},
  {Falcke}, {Moscibrodzka}, {O'Leary}, {Eatough}, {Kramer}, {Lee}, {Spitler},
  {Desvignes}, {Rushton}, {Doeleman}, \& {Reid}}]{2015ApJ...798..120B}
{Bower}, G.~C., {Deller}, A., {Demorest}, P., {et~al.} 2015, \apj, 798, 120,
  \dodoi{10.1088/0004-637X/798/2/120}

\bibitem[{{Deller} {et~al.}(2007){Deller}, {Tingay}, {Bailes}, \&
  {West}}]{2007PASP..119..318D}
{Deller}, A.~T., {Tingay}, S.~J., {Bailes}, M., \& {West}, C. 2007, \pasp, 119,
  318, \dodoi{10.1086/513572}

\bibitem[{{Event Horizon Telescope Collaboration} {et~al.}(2022){Event Horizon
  Telescope Collaboration}, {Akiyama}, {Alberdi}, {Alef}, {Algaba}, {Anantua},
  {Asada}, {Azulay}, {Bach}, {Baczko}, {Ball}, {Balokovi{\'c}}, {Barrett},
  {Baub{\"o}ck}, {Benson}, {Bintley}, {Blackburn}, {Blundell}, {Bouman},
  {Bower}, {Boyce}, {Bremer}, {Brinkerink}, {Brissenden}, {Britzen},
  {Broderick}, {Broguiere}, {Bronzwaer}, {Bustamante}, {Byun}, {Carlstrom},
  {Ceccobello}, {Chael}, {Chan}, {Chatterjee}, {Chatterjee}, {Chen}, {Chen},
  {Cheng}, {Cho}, {Christian}, {Conroy}, {Conway}, {Cordes}, {Crawford},
  {Crew}, {Cruz-Osorio}, {Cui}, {Davelaar}, {Laurentis}, {Deane}, {Dempsey},
  {Desvignes}, {Dexter}, {Dhruv}, {Doeleman}, {Dougal}, {Dzib}, {Eatough},
  {Emami}, {Falcke}, {Farah}, {Fish}, {Fomalont}, {Ford}, {Fraga-Encinas},
  {Freeman}, {Friberg}, {Fromm}, {Fuentes}, {Galison}, {Gammie}, {Garc{\'\i}a},
  {Gentaz}, {Georgiev}, {Goddi}, {Gold}, {G{\'o}mez-Ruiz}, {G{\'o}mez}, {Gu},
  {Gurwell}, {Hada}, {Haggard}, {Haworth}, {Hecht}, {Hesper}, {Heumann}, {Ho},
  {Ho}, {Honma}, {Huang}, {Huang}, {Hughes}, {Ikeda}, {Impellizzeri}, {Inoue},
  {Issaoun}, {James}, {Jannuzi}, {Janssen}, {Jeter}, {Jiang},
  {Jim{\'e}nez-Rosales}, {Johnson}, {Jorstad}, {Joshi}, {Jung}, {Karami},
  {Karuppusamy}, {Kawashima}, {Keating}, {Kettenis}, {Kim}, {Kim}, {Kim},
  {Kim}, {Kino}, {Koay}, {Kocherlakota}, {Kofuji}, {Koch}, {Koyama}, {Kramer},
  {Kramer}, {Krichbaum}, {Kuo}, {Bella}, {Lauer}, {Lee}, {Lee}, {Leung},
  {Levis}, {Li}, {Lico}, {Lindahl}, {Lindqvist}, {Lisakov}, {Liu}, {Liu},
  {Liuzzo}, {Lo}, {Lobanov}, {Loinard}, {Lonsdale}, {Lu}, {Mao}, {Marchili},
  {Markoff}, {Marrone}, {Marscher}, {Mart{\'\i}-Vidal}, {Matsushita},
  {Matthews}, {Medeiros}, {Menten}, {Michalik}, {Mizuno}, {Mizuno}, {Moran},
  {Moriyama}, {Moscibrodzka}, {M{\"u}ller}, {Mus}, {Musoke}, {Myserlis},
  {Nadolski}, {Nagai}, {Nagar}, {Nakamura}, {Narayan}, {Narayanan},
  {Natarajan}, {Nathanail}, {Fuentes}, {Neilsen}, {Neri}, {Ni}, {Noutsos},
  {Nowak}, {Oh}, {Okino}, {Olivares}, {Ortiz-Le{\'o}n}, {Oyama}, {{\"O}zel},
  {Palumbo}, {Paraschos}, {Park}, {Parsons}, {Patel}, {Pen}, {Pesce},
  {Pi{\'e}tu}, {Plambeck}, {PopStefanija}, {Porth}, {P{\"o}tzl}, {Prather},
  {Preciado-L{\'o}pez}, {Psaltis}, {Pu}, {Ramakrishnan}, {Rao}, {Rawlings},
  {Raymond}, {Rezzolla}, {Ricarte}, {Ripperda}, {Roelofs}, {Rogers}, {Ros},
  {Romero-Ca{\~n}izales}, {Roshanineshat}, {Rottmann}, {Roy}, {Ruiz},
  {Ruszczyk}, {Rygl}, {S{\'a}nchez}, {S{\'a}nchez-Arg{\"u}elles},
  {S{\'a}nchez-Portal}, {Sasada}, {Satapathy}, {Savolainen}, {Schloerb},
  {Schonfeld}, {Schuster}, {Shao}, {Shen}, {Small}, {Sohn}, {SooHoo},
  {Souccar}, {Sun}, {Tazaki}, {Tetarenko}, {Tiede}, {Tilanus}, {Titus},
  {Torne}, {Traianou}, {Trent}, {Trippe}, {Turk}, {van Bemmel}, {van
  Langevelde}, {van Rossum}, {Vos}, {Wagner}, {Ward-Thompson}, {Wardle},
  {Weintroub}, {Wex}, {Wharton}, {Wielgus}, {Wiik}, {Witzel}, {Wondrak},
  {Wong}, {Wu}, {Yamaguchi}, {Yoon}, {Young}, {Young}, {Younsi}, {Yuan},
  {Yuan}, {Zensus}, {Zhang}, {Zhao}, {Zhao}, {Agurto}, {Allardi}, {Amestica},
  {Araneda}, {Arriagada}, {Berghuis}, {Bertarini}, {Berthold}, {Blanchard},
  {Brown}, {C{\'a}rdenas}, {Cantzler}, {Caro}, {Castillo-Dom{\'\i}nguez},
  {Chan}, {Chang}, {Chang}, {Chang}, {Chang}, {Chen}, {Chilson}, {Chuter},
  {Ciechanowicz}, {Colin-Beltran}, {Coulson}, {Crowley}, {Degenaar},
  {Dornbusch}, {Dur{\'a}n}, {Everett}, {Faber}, {Forster}, {Fuchs}, {Gale},
  {Geertsema}, {Gonz{\'a}lez}, {Graham}, {Gueth}, {Halverson}, {Han}, {Han},
  {Hasegawa}, {Hern{\'a}ndez-Rebollar}, {Herrera}, {Herrero-Illana},
  {Heyminck}, {Hirota}, {Hoge}, {Hostler Schimpf}, {Howie}, {Huang}, {Jiang},
  {Jinchi}, {John}, {Kimura}, {Klein}, {Kubo}, {Kuroda}, {Kwon}, {Lacasse},
  {Laing}, {Leitch}, {Li}, {Liu}, {Liu}, {Lin}, {Lu}, {Mac-Auliffe},
  {Martin-Cocher}, {Matulonis}, {Maute}, {Messias}, {Meyer-Zhao},
  {Monta{\~n}a}, {Montenegro-Montes}, {Montgomerie}, {Moreno Nolasco},
  {Muders}, {Nishioka}, {Norton}, {Nystrom}, {Ogawa}, {Olivares}, {Oshiro},
  {P{\'e}rez-Beaupuits}, {Parra}, {Phillips}, {Poirier}, {Pradel}, {Qiu},
  {Raffin}, {Rahlin}, {Ram{\'\i}rez}, {Ressler}, {Reynolds},
  {Rodr{\'\i}guez-Montoya}, {Saez-Madain}, {Santana}, {Shaw}, {Shirkey},
  {Silva}, {Snow}, {Sousa}, {Sridharan}, {Stahm}, {Stark}, {Test},
  {Torstensson}, {Venegas}, {Walther}, {Wei}, {White}, {Wieching}, {Wijnands},
  {Wouterloot}, {Yu}, {Yu (于威)}, \& {Zeballos}}]{2022ApJ...930L..12E}
{Event Horizon Telescope Collaboration}, {Akiyama}, K., {Alberdi}, A., {et~al.}
  2022, \apjl, 930, L12, \dodoi{10.3847/2041-8213/ac6674}

\bibitem[{{Genzel} {et~al.}(2003){Genzel}, {Sch{\"o}del}, {Ott}, {Eckart},
  {Alexander}, {Lacombe}, {Rouan}, \& {Aschenbach}}]{2003Natur.425..934G}
{Genzel}, R., {Sch{\"o}del}, R., {Ott}, T., {et~al.} 2003, \nat, 425, 934,
  \dodoi{10.1038/nature02065}

\bibitem[{{Ghez} {et~al.}(2008){Ghez}, {Salim}, {Weinberg}, {Lu}, {Do}, {Dunn},
  {Matthews}, {Morris}, {Yelda}, {Becklin}, {Kremenek}, {Milosavljevic}, \&
  {Naiman}}]{2008ApJ...689.1044G}
{Ghez}, A.~M., {Salim}, S., {Weinberg}, N.~N., {et~al.} 2008, \apj, 689, 1044,
  \dodoi{10.1086/592738}

\bibitem[{{Gillessen} {et~al.}(2009){Gillessen}, {Eisenhauer}, {Trippe},
  {Alexander}, {Genzel}, {Martins}, \& {Ott}}]{2009ApJ...692.1075G}
{Gillessen}, S., {Eisenhauer}, F., {Trippe}, S., {et~al.} 2009, \apj, 692,
  1075, \dodoi{10.1088/0004-637X/692/2/1075}

\bibitem[{{Gordon} {et~al.}(2022){Gordon}, {Jacobs}, \& {de
  Witt}}]{2022ivs..meetE...1G}
{Gordon}, D., {Jacobs}, C.~S., \& {de Witt}, A. 2022, in Proc. 12th IVS General
  Meeting, 1

\bibitem[{{Gravity Collaboration} {et~al.}(2017){Gravity Collaboration},
  {Abuter}, {Accardo}, {Amorim}, {Anugu}, {{\'A}vila}, {Azouaoui}, {Benisty},
  {Berger}, {Blind}, {Bonnet}, {Bourget}, {Brandner}, {Brast}, {Buron},
  {Burtscher}, {Cassaing}, {Chapron}, {Choquet}, {Cl{\'e}net}, {Collin},
  {Coud{\'e} Du Foresto}, {de Wit}, {de Zeeuw}, {Deen},
  {Delplancke-Str{\"o}bele}, {Dembet}, {Derie}, {Dexter}, {Duvert}, {Ebert},
  {Eckart}, {Eisenhauer}, {Esselborn}, {F{\'e}dou}, {Finger}, {Garcia}, {Garcia
  Dabo}, {Garcia Lopez}, {Gendron}, {Genzel}, {Gillessen}, {Gonte}, {Gordo},
  {Grould}, {Gr{\"o}zinger}, {Guieu}, {Haguenauer}, {Hans}, {Haubois}, {Haug},
  {Haussmann}, {Henning}, {Hippler}, {Horrobin}, {Huber}, {Hubert}, {Hubin},
  {Hummel}, {Jakob}, {Janssen}, {Jochum}, {Jocou}, {Kaufer}, {Kellner},
  {Kendrew}, {Kern}, {Kervella}, {Kiekebusch}, {Klein}, {Kok}, {Kolb}, {Kulas},
  {Lacour}, {Lapeyr{\`e}re}, {Lazareff}, {Le Bouquin}, {L{\`e}na}, {Lenzen},
  {L{\'e}v{\^e}que}, {Lippa}, {Magnard}, {Mehrgan}, {Mellein}, {M{\'e}rand},
  {Moreno-Ventas}, {Moulin}, {M{\"u}ller}, {M{\"u}ller}, {Neumann}, {Oberti},
  {Ott}, {Pallanca}, {Panduro}, {Pasquini}, {Paumard}, {Percheron}, {Perraut},
  {Perrin}, {Pfl{\"u}ger}, {Pfuhl}, {Phan Duc}, {Plewa}, {Popovic}, {Rabien},
  {Ram{\'\i}rez}, {Ramos}, {Rau}, {Riquelme}, {Rohloff}, {Rousset},
  {Sanchez-Bermudez}, {Scheithauer}, {Sch{\"o}ller}, {Schuhler}, {Spyromilio},
  {Straubmeier}, {Sturm}, {Suarez}, {Tristram}, {Ventura}, {Vincent},
  {Waisberg}, {Wank}, {Weber}, {Wieprecht}, {Wiest}, {Wiezorrek}, {Wittkowski},
  {Woillez}, {Wolff}, {Yazici}, {Ziegler}, \& {Zins}}]{2017A&A...602A..94G}
{Gravity Collaboration}, {Abuter}, R., {Accardo}, M., {et~al.} 2017, \aap, 602,
  A94, \dodoi{10.1051/0004-6361/201730838}

\bibitem[{{Gravity Collaboration} {et~al.}(2019){Gravity Collaboration},
  {Abuter}, {Amorim}, {Baub{\"o}ck}, {Berger}, {Bonnet}, {Brandner},
  {Cl{\'e}net}, {Coud{\'e} Du Foresto}, {de Zeeuw}, {Dexter}, {Duvert},
  {Eckart}, {Eisenhauer}, {F{\"o}rster Schreiber}, {Garcia}, {Gao}, {Gendron},
  {Genzel}, {Gerhard}, {Gillessen}, {Habibi}, {Haubois}, {Henning}, {Hippler},
  {Horrobin}, {Jim{\'e}nez-Rosales}, {Jocou}, {Kervella}, {Lacour},
  {Lapeyr{\`e}re}, {Le Bouquin}, {L{\'e}na}, {Ott}, {Paumard}, {Perraut},
  {Perrin}, {Pfuhl}, {Rabien}, {Rodriguez Coira}, {Rousset}, {Scheithauer},
  {Sternberg}, {Straub}, {Straubmeier}, {Sturm}, {Tacconi}, {Vincent}, {von
  Fellenberg}, {Waisberg}, {Widmann}, {Wieprecht}, {Wiezorrek}, {Woillez}, \&
  {Yazici}}]{2019A&A...625L..10G}
{Gravity Collaboration}, {Abuter}, R., {Amorim}, A., {et~al.} 2019, \aap, 625,
  L10, \dodoi{10.1051/0004-6361/201935656}

\bibitem[{{Greisen}(2003)}]{2003ASSL..285..109G}
{Greisen}, E.~W. 2003, in Astrophysics and Space Science Library, Vol. 285,
  Information Handling in Astronomy - Historical Vistas, ed. A.~{Heck}, 109,
  \dodoi{10.1007/0-306-48080-8\_7}

\bibitem[{{Hada} {et~al.}(2011){Hada}, {Doi}, {Kino}, {Nagai}, {Hagiwara}, \&
  {Kawaguchi}}]{2011Natur.477..185H}
{Hada}, K., {Doi}, A., {Kino}, M., {et~al.} 2011, \nat, 477, 185,
  \dodoi{10.1038/nature10387}

\bibitem[{{Lane}(1979)}]{1979PASP...91..405L}
{Lane}, A.~P. 1979, \pasp, 91, 405, \dodoi{10.1086/130508}

\bibitem[{{Lo} {et~al.}(1998){Lo}, {Shen}, {Zhao}, \&
  {Ho}}]{1998ApJ...508L..61L}
{Lo}, K.~Y., {Shen}, Z.-Q., {Zhao}, J.-H., \& {Ho}, P. T.~P. 1998, \apjl, 508,
  L61, \dodoi{10.1086/311726}

\bibitem[{{Ma} {et~al.}(1998){Ma}, {Arias}, {Eubanks}, {Fey}, {Gontier},
  {Jacobs}, {Sovers}, {Archinal}, \& {Charlot}}]{1998AJ....116..516M}
{Ma}, C., {Arias}, E.~F., {Eubanks}, T.~M., {et~al.} 1998, \aj, 116, 516,
  \dodoi{10.1086/300408}

\bibitem[{{Marcaide} {et~al.}(1992){Marcaide}, {Alberdi}, {Bartel}, {Clark},
  {Corey}, {Elosegui}, {Gorenstein}, {Guirado}, {Kardashev}, {Popov},
  {Preston}, {Ratner}, {Rioja}, {Rogers}, \& {Shapiro}}]{1992A&A...258..295M}
{Marcaide}, J.~M., {Alberdi}, A., {Bartel}, N., {et~al.} 1992, \aap, 258, 295

\bibitem[{{Menten} {et~al.}(1997){Menten}, {Reid}, {Eckart}, \&
  {Genzel}}]{1997ApJ...475L.111M}
{Menten}, K.~M., {Reid}, M.~J., {Eckart}, A., \& {Genzel}, R. 1997, \apjl, 475,
  L111, \dodoi{10.1086/310472}

\bibitem[{{Pradel} {et~al.}(2006){Pradel}, {Charlot}, \&
  {Lestrade}}]{2006A&A...452.1099P}
{Pradel}, N., {Charlot}, P., \& {Lestrade}, J.-F. 2006, \aap, 452, 1099,
  \dodoi{10.1051/0004-6361:20053021}

\bibitem[{{Reid} \& {Brunthaler}(2004)}]{2004ApJ...616..872R}
{Reid}, M.~J., \& {Brunthaler}, A. 2004, \apj, 616, 872, \dodoi{10.1086/424960}

\bibitem[{{Reid} \& {Brunthaler}(2020)}]{2020ApJ...892...39R}
---. 2020, \apj, 892, 39, \dodoi{10.3847/1538-4357/ab76cd}

\bibitem[{{Reid} \& {Honma}(2014)}]{2014ARA&A..52..339R}
{Reid}, M.~J., \& {Honma}, M. 2014, \araa, 52, 339,
  \dodoi{10.1146/annurev-astro-081913-040006}

\bibitem[{{Reid} {et~al.}(2003){Reid}, {Menten}, {Genzel}, {Ott},
  {Sch{\"o}del}, \& {Eckart}}]{2003ApJ...587..208R}
{Reid}, M.~J., {Menten}, K.~M., {Genzel}, R., {et~al.} 2003, \apj, 587, 208,
  \dodoi{10.1086/368074}

\bibitem[{{Reid} {et~al.}(2019){Reid}, {Menten}, {Brunthaler}, {Zheng}, {Dame},
  {Xu}, {Li}, {Sakai}, {Wu}, {Immer}, {Zhang}, {Sanna}, {Moscadelli}, {Rygl},
  {Bartkiewicz}, {Hu}, {Quiroga-Nu{\~n}ez}, \& {van
  Langevelde}}]{2019ApJ...885..131R}
{Reid}, M.~J., {Menten}, K.~M., {Brunthaler}, A., {et~al.} 2019, \apj, 885,
  131, \dodoi{10.3847/1538-4357/ab4a11}

\bibitem[{{Rioja} \& {Dodson}(2020)}]{2020A&ARv..28....6R}
{Rioja}, M.~J., \& {Dodson}, R. 2020, \aapr, 28, 6,
  \dodoi{10.1007/s00159-020-00126-z}

\bibitem[{{Rogers} {et~al.}(1994){Rogers}, {Doeleman}, {Wright}, {Bower},
  {Backer}, {Padin}, {Philips}, {Emerson}, {Greenhill}, {Moran}, \&
  {Kellermann}}]{1994ApJ...434L..59R}
{Rogers}, A. E.~E., {Doeleman}, S., {Wright}, M. C.~H., {et~al.} 1994, \apjl,
  434, L59, \dodoi{10.1086/187574}

\bibitem[{{Sakai} {et~al.}(2019){Sakai}, {Lu}, {Ghez}, {Jia}, {Do}, {Witzel},
  {Gautam}, {Hees}, {Becklin}, {Matthews}, \& {Hosek}}]{2019ApJ...873...65S}
{Sakai}, S., {Lu}, J.~R., {Ghez}, A., {et~al.} 2019, \apj, 873, 65,
  \dodoi{10.3847/1538-4357/ab0361}

\bibitem[{{Scoville} {et~al.}(2003){Scoville}, {Stolovy}, {Rieke},
  {Christopher}, \& {Yusef-Zadeh}}]{2003ApJ...594..294S}
{Scoville}, N.~Z., {Stolovy}, S.~R., {Rieke}, M., {Christopher}, M., \&
  {Yusef-Zadeh}, F. 2003, \apj, 594, 294, \dodoi{10.1086/376790}

\bibitem[{{Thompson} {et~al.}(2017){Thompson}, {Moran}, \&
  {Swenson}}]{2017isra.book.....T}
{Thompson}, A.~R., {Moran}, J.~M., \& {Swenson}, George~W., J. 2017,
  {Interferometry and Synthesis in Radio Astronomy, 3rd Edition},
  \dodoi{10.1007/978-3-319-44431-4}

\bibitem[{{Tsuboi} {et~al.}(2022){Tsuboi}, {Tsutsumi}, {Miyazaki}, {Miyawaki},
  \& {Miyoshi}}]{2022arXiv220406778T}
{Tsuboi}, M., {Tsutsumi}, T., {Miyazaki}, A., {Miyawaki}, R., \& {Miyoshi}, M.
  2022, arXiv e-prints, arXiv:2204.06778.
\newblock \doarXiv{2204.06778}

\bibitem[{{Yusef-Zadeh} {et~al.}(1999){Yusef-Zadeh}, {Choate}, \&
  {Cotton}}]{1999ApJ...518L..33Y}
{Yusef-Zadeh}, F., {Choate}, D., \& {Cotton}, W. 1999, \apjl, 518, L33,
  \dodoi{10.1086/312058}

\end{thebibliography}
 
\end{CJK*} 
\end{document}